\documentclass[showpacs,preprintnumbers,amsmath,amssymb]{revtex4}
\makeatletter
\begin{document}
%input{epsf}
%\maketitle
\title{ Spin Entanglement of Two delocalised  Fermions and Berry Phase}
\author{B. Basu}
 \email{banasri@isical.ac.in}
\author{P. Bandyopadhyay}
 \email{b_pratul@yahoo.co.in}
 \affiliation{Physics and Applied Mathematics Unit\\
 Indian Statistical Institute\\
 Kolkata-700108 }

\begin{abstract}
 We have studied the entanglement of identical fermions in two spatial regions in terms of the Berry phase acquired by their spins. The analysis is done from the viewpoint of the geometrical interpretation of entanglement, where a fermion is visualized as a scalar particle attached with a magnetic flux quantum. The quantification of spin entanglement in
 terms of their  Berry phases is novel and generalises the relationship between the entanglement of distinguishable spins and that of delocalised fermions.
  \end{abstract}
\maketitle
\section{Introduction}
Quantum entanglement is a specific feature which distinguishes  between the classical and quantum world. The role of entanglement
is also important in different branches of quantum information science
such as quantum communication\cite{sch}, quantum computation
\cite{sc}, quantum cryptography \cite{ek} and quantum teleportation
\cite{ben}. Entanglement for two distinguishable qubits have been
well studied and a measure of the degree of entanglement can be
quantified in terms of von Neuman entropy and concurrence
\cite{a1,b1,c1,d1}. However, entanglement of two 
identical fermions have not yet been well understood.  In systems of identical fermions, a proper measure of entanglement should take into account multiple occupancy of states \cite{3,4,5,6,7}, the effect of exchange \cite{2} and mutual repulsion. Recently,
Ramsak et al. \cite{8} have considered the problem  and formulated
several expressions for the concurrence of two indistinguishable
delocalised spin $1/2$ particles. In a recent paper \cite{1}, it has
been pointed out that the concurrence for the entanglement of two
distinguishable spins can be formulated in terms of the Berry phase
acquired by the spins when each spin is rotated about the
quantization axis(z-axis). In fact, when a spinor is visualized as a
scalar particle attached with a magnetic flux, quantum entanglement
of spin systems is caused by the deviation of the internal magnetic
flux line associated with one particle in presence of the other.
This helps us to consider the measure of entanglement viz.
concurrence, in terms of the Berry phase acquired by the rotation of
the spin around the z-axis induced by the
 internal magnetic field of the other particle.
 This picture is potentially useful to study
the entanglement of identical fermions in two spatial regions in terms of the Berry phase
acquired by their spins. Indeed in this formalism, the spin entanglement through magnetic coupling is associated with
 the spatial entanglement between fermions at different spatial regions and entanglement
 can be viewed as a consequence of Fermi statistics \cite{2}
Therefore, just like in distinguishable spin systems, the concurrence associated with the entanglement of identical fermions in different spatial regions can also be expressed in terms of the geometrical  phase. The phase is  acquired by the spin of one particle in one spatial region, when it moves around the z-axis in presence of the other particle, in another spatial region.
In the present note, we shall study the entanglement of two delocalised electrons in two spatial regions from this viewpoint.

\section{Concurrence and Berry Phase}
For an entangled state, the Berry phase acquired by a spin may be analysed by considering
that, under the influence of the internal magnetic field   associated with the other
electron, the spin of an electron rotates adiabatically with an angular velocity
$\omega_0$ around the $z$-axis under an angle $\theta$.

 The instantaneous eigenstates of a spin operator in direction
 ${\bf n}(\theta,t)$ where ${\bf n}$ is the unit vector depicting the magnetic field
  ${\bf B}(t)=B {\bf n}(\theta,t)$ in the $\sigma_z$-basis are given by
  \begin{equation}\label{art}
\begin{array}{ccc}
    \displaystyle{|\uparrow_n;t>}&=&\displaystyle{\cos \frac{\theta}{2} |\uparrow_z> +~ \sin
    \frac{\theta}{2} e^{i\omega_0 t}|\downarrow_z> }\\
    &&\\
    \displaystyle{|\downarrow_n;t>}&=&\displaystyle{\sin \frac{\theta}{2} |\uparrow_z> +~ \cos
    \frac{\theta}{2} e^{i\omega_0 t}|\downarrow_z>}
\end{array}
\end{equation}

After cyclic evolution for the interval
$\tau=\displaystyle{\frac{2\pi}{\omega_0}}$ each eigenstate will
pick up a geometric phase (Berry phase) apart from the dynamical
phase \cite{10}
\begin{equation}\label{a16}
      \displaystyle{\Phi_{B \mp}}~=  \displaystyle{\pi(1 \mp \cos \theta)}
    \end{equation}
    where $\Phi_{B-}(\Phi_{B+})$ corresponds to up (down) state.
     The angle $\theta$ represents the deviation of the spin from the quantization axis (z-axis)under the influence of the magnetic field.

The evaluation of the concurrence in terms of the Berry phase follows from the following consideration. For the Bell state
\begin{equation}\label{Bel}
|\psi>=a |\uparrow\downarrow>-b |\downarrow\uparrow>
\end{equation}
where $a$ and $b$ are complex coefficients, the concurrence is given by
\begin{equation}
C=2|a||b|
\end{equation}

 In this formalism, as entanglement is considered to be caused by the deviation of the magnetic flux line from the quantization axis in presence of the other particle, we may take $|a|$ and $|b|$ as functions of this angle of deviation $\theta$ and thus we write
 \begin{equation}
\frac{1}{\sqrt 2}\left(
\begin{array}{c}
|a|\\
|b|
\end{array}
\right) =\left(
\begin{array}{c}
f(\theta)\\
g(\theta)
\end{array}
\right)
\end{equation}
The angle $\theta$ here just corresponds to the deviation of up(down) spin under the influence of the other and thus represents the same angle $\theta$ associated with the Berry phase acquired by the spin as given by eqn. (2).
For the maximum entangled state (MES), we have $\theta=\pi$  as it corresponds to the maximum deviation of a spin from the z-axis when the spin direction is reversed. For this state, we have $ \mid a\mid =~\mid b\mid~=\frac{1}{\sqrt 2} $ and  $C=1$. \\
Again for the disentangled state $\theta=0 $ and we have $C=0$\\
These constraints satisfy,
\begin{equation}
f(\theta)\mid_{\theta=\pi}=~g(\theta)\mid_{\theta=\pi}=\frac{1}{2}
\end{equation}
and,
\begin{equation}
\rm{either}~f(\theta)\mid_{\theta=0}=0~~~~~~~~\rm{or}~g(\theta)\mid_{\theta=0}=0
\end{equation}
From these constraint equations, for the positive definite norms
$0\leq \mid a \mid \leq 1$ and $0\leq \mid b \mid \leq 1$,
we can have a general solution
\begin{equation}
\frac{1}{\sqrt{2}}\left(
\begin{array}{c}
|a|\\
|b|
\end{array}
\right) =
\left(
\begin{array}{c}
f(\theta)\\
g(\theta)
\end{array}
\right)=
\left(
\begin{array}{c}
\cos^2\frac{n\theta}{4}\\
\sin^2\frac{n\theta}{4}
\end{array}
\right)
\end{equation}
with $n$ being an odd integer.
It is noted that according to eqn. (8) the relation $|a|^2 +|b|^2=1$ is satisfied only in the case of $\theta=\pi$ implying the MES. So to have the probability interpretation the generalised state may be defined by incorporating the normalization factor $\frac{1}{\sqrt{|a|^2 +|b|^2}}$ in eqn. (3). The Berry phase corresponds to the half of the solid angle $\frac{1}{2}\Omega$
swept out by the magnetic flux line and is given by $\pi(1-\cos\theta)$. The system under consideration suggests that the range of $\theta$ lies between
$0\leq \mid \theta \mid \leq \pi$ where $\theta =\pi$ corresponds to the maximum deviation of the spin when the spin direction is reversed. So in the expression (8) we should take
$n=1$ for our present system. We find that the particular solution with $n=1$ relates the concurrence with the Berry phase and is given by
 \begin{equation}
C=2|a|~|b|=\sin^2\frac{\theta}{2}=\frac{1}{2}(1-\cos\theta) = \frac{|\phi_B|}{2\pi} \end{equation}

We may remark here that the concurrence (as it is a measure of entanglement) is a function of an instantaneous state,  whereas the Berry phase is related to the periodic rotation of the system. The relationship between these two entities in the present framework follows from physical aspects. Here, entanglement is caused by the deviation of the magnetic flux line associated with one fermion in  presence of the other and the Berry phase of an entangled spin system is related with this deviation. This is the novelty of  studying spin entanglement from Berry phase approach.

\section{Spin Entanglement of Two delocalised Fermions}
 In our framework, we consider  two
electrons in two different spatial regions A and B. Entanglement is
produced when two initially unentangled(separated) electrons in wave
packets approach each other, interact and then again become well
separated into distinct regions A and B. The spin properties of such
a fermionic system  can be realized in spin correlation functions
for the two domains. In fact, the spin measuring apparatus could
measure spin correlation functions for the two domains A and B
rather than two distinguishable spins. We may consider spin entanglement
of two-electron states on a lattice of the form
\begin{equation}
|\psi>=\sum_{i,j=1}^N \frac{1}{2}\left[ \psi_{i j}^{\uparrow \downarrow}
c_{i \uparrow}^\dagger  c_{j \downarrow}^\dagger~+~
\psi_{i j}^{\downarrow \uparrow}
c_{i \downarrow}^\dagger  c_{j \uparrow}^\dagger~\right] |0>
\end{equation}
where $c_{i s}^\dagger$ creates an electron with spin $s$ on site $i$ and $N$ is the total number of sites.
Here $\psi_{i j}^{\uparrow \downarrow} (\psi_{i j}^{\downarrow \uparrow})$ is the amplitude of probability to find the two-electron state with one having spin $\uparrow$ in region A and another with spin $\downarrow$ in region B. The whole set of probabilities give the wave function for the two-electron system in the continuum limit. 

The system is relevant in representing a tight binding lattice containing two valence electrons occupying two non-degenerate atomic orbitals or two electrons in the conduction band of a semiconductor for which the site represents finite grid points.

To study the concurrence associated with the entanglement of such a system in terms of the $geometric$ $phase$ acquired by the spin of one electron in presence of the other electron,  we consider a rotation of the spin around the $z$-axis under an angle $\theta$ at each site
\begin{equation}
 \psi_{i j}^{\uparrow \downarrow} \rightarrow \psi_{i j}^{\uparrow \downarrow}e^{2i\theta}
 \end{equation}
 when the angle $\theta$ varies from 0 to $\pi$. The Berry phase acquired by the spin may be realised through the expression
 \begin{equation}
 \Phi_B=-i~\int_0^\pi ~<\psi|\partial_\theta \psi>d\theta
 \end{equation}
 which on the lattice takes the form
 \begin{equation}\label{a4}
 \Phi_B=2\pi ~2 \sum_{i,j}\psi_{i j}^{\uparrow \downarrow ^{*}}~\psi_{j i}^{\uparrow \downarrow}
 \end{equation}
This follows from the differentiation of the expression (11) with respect to $\theta$ and replacing the integration in the continuum case by the summation on the lattice.
 The relationship between concurrence and the Berry phase can be generalised for the system of two indistinguishable particles and from eqns. (9) and (\ref{a4}) we can write
 \begin{equation}\label{u1}
 C=\frac{|\Phi_B|}{2\pi}=2 \sum_{i,j}\psi_{i j}^{\uparrow \downarrow ^{*}}~\psi_{j i}^{\uparrow \downarrow}
 \end{equation}
 This may be identified with the formula obtained by Ramsak et. al.\cite{8} for the entanglement of the two electron states on a lattice (given by eqn. (10)). The concurrence of the system can be expressed in terms of the operators
 \begin{equation}
 S^+_{A(B)}=(S^-_{A(B)})^\dagger=\sum_{i\in A(B)}c^\dagger_{i \uparrow}c_{i \downarrow}
 \end{equation}
 and for the state with $S^Z_{tot}=0$, we have
 \begin{equation}\label{r1}
 C=2|<S^+_A~ S^-_B>|~=~2\sum_{i,j}\psi_{i j}^{\uparrow \downarrow ^{*}}~\psi_{j i}^{\uparrow \downarrow}
 \end{equation}
 Indeed, this can be formulated in a more familiar form by considering the state in analogy to the Bell state
 \begin{equation}
 \Phi^{\pm}_{i j}=\frac{1}{\sqrt 2}(\psi_{i j}^{\uparrow \downarrow} \pm
 \psi_{j i}^{\uparrow \downarrow} )
 \end{equation}
 over all pairs $[i j]$ such that $i\in A$ and $j \in B$. The expression for concurrence of the system is given by \cite{8}
 \begin{equation}
 C= \sum_{[i j]}\mid [(\Phi_{i j}^+)^2~-~(\Phi_{i j}^-)^2 ]\mid
 \end{equation}
 which is equivalent to the expression (\ref{r1}). From our analysis we note that this result is identical with the expression (\ref{u1}) obtained from the relationship of Berry phase with concurrence.

 \section{Entanglement of two delocalised electrons in Hubbard model }
 As the study of generation of entanglement in the solid state environment is an active field of research in recent times, for an application for our formalism we have picked up the well studied Hubbard model  \cite{11}.
  
 To compute the concurrence for the entanglement of two electrons in two different spatial regions in Hubbard model, let us consider two interacting electrons in a one dimensional lattice with $N \rightarrow \infty $. The corresponding Hamiltonian is
 \begin{equation}\label{a6}
 H=-t~\sum _{i j}\left(c^\dagger _{i~ s}c_{j~ s}+h.c.\right)+
   \sum _{i, j, s, s^\prime}U_{i`j}n_{i~ s}n_{j~ s^\prime}
\end{equation}
where $t$ is the hopping parameter, $U$ represents the onsite repulsion and $n_{i~ s}$ is the number of electrons at the site $i$ with spin $s$.
Let the situation be such, that one electron with spin ${\uparrow}$
is initially confined in the region A and the other electron with
opposite spin  ${\downarrow}$ in region B. The initial state is
defined by two wave packets, the left with momentum $k$ and the
right with momentum $-q$. After collision, the electrons move apart
with non-spinflip  amplitude $t_{kq}$ and spin flip amplitude
$r_{kq}$. For sharp momentum resolutions we take $k=-q=k_0$.
 We would like to study the entanglement of these two electrons in terms of the Berry phase acquired by the spins in this system.
We know that for strong coupling and at half filling, the system with Hamiltonian (\ref{a6}) reduces to  the Heisenberg antiferromagnetic chain and the Hamiltonain is given by
  \begin{equation}
  H=J\sum \left[ S^x_i~ S^x_j~+~S^y_i~ S^y_j~+~S^z_i~ S^z_j\right]
  \end{equation}
  with $J=4t^2/U$. In the $S=0$ sector( $S$ = total spin), the rotational symmetry of the Hamiltonian implies
  \begin{equation}
  <S^x_i~ S^x_j>~=~<S^y_i~ S^y_j>~=~<S^z_i~ S^z_j >
  \end{equation}
  In the antiferromagnetic chain for spin 1/2 system,
  \begin{equation}
  <S^z_i~ S^z_i>~\leq \frac{1}{4}
  \end{equation}
  If $\theta$ be the deviation of the spin at the site $i$ from the  quantization axis  i.e. $z$-axis under the influence of the spin at the site $j$ then we can write,
  \begin{equation}
  <S^z_i~ S^z_j>=\frac{1}{4}\cos \theta
  \end{equation}
We consider  collision of the two electrons initially at the regions
A and B. After the collision the electrons move to the final states
in these two regions either with spin flip or non-spin flip
configurations. The Berry phase acquired by the up(down)
configuration is given by
  $$\Phi_{B-}(\Phi_{B+})~= \pi(1 - \cos \theta)(\pi(1 + \cos \theta))$$
    However after the collision, the initial spin positions get changed so that for spin flip and spin nonflip cases we have the two phases
    \begin{equation}
    \Phi_B= \pi(1 - \cos \theta)|_{\theta=\pi}~~~~\rm{and}~~~~  \Phi_B=\pi(1 + \cos \theta)|_{\theta=0}
    \end{equation}
    respectively.

   % \hspace{4cm}

   The generalised  expression for the Berry phase is
     \begin{equation}
    \Phi_B= \pi(1 + |\cos \theta)|)
    \end{equation}
    When the spin flip and spin nonflip amplitudes coincide
     %$t_{kq}=r_{kq}$
    the concurrence is given by
   \begin{equation}
    C=\frac{|\Phi_B|}{2\pi}=\frac{1}{2}(1 + |\cos \theta)|)|_{\theta=0,\pi}=1
    \end{equation}
    Our result is identical with another definition of concurrence \cite{8}
\begin{equation}
C=2|t_{kq}r_{kq}|=1
\end{equation}
when the spin flip and spin nonflip amplitude coincides i.e.
     $t_{kq}=r_{kq}$
      This corresponds to $k_0 \sim 0,\pi$. However, when the spin flip and non-spin-flip amplitudes do not coincide i.e. $t_{kq}\neq r_{kq}$, we can measure the concurrence
    from an estimate of the angle $\theta$ in terms of momentum $k_0(k=-q=k_0)$. This can be achieved from an analysis of the energy relations in Hubbard model and Heisenberg antiferromagentic chain in the ground state with site $i \in A, j\in B$ .
    In Hubbard model, when no particles meet at a lattice point, the many particle energy is given by
    \begin{equation}
    E=-2t \sum_i \cos k_i
    \end{equation}
    In the Heisenberg antiferromagnetic chain with the correlation given by eqn.(23),
 the energy per site is given by
 \begin{equation}
    E=J \frac{3}{4} \cos \theta
    \end{equation}
 %Taking into account that the Hubbard model reduces to the Heisenberg antiferromagnetic chain at half filling
Since in the Hubbard model, the occupation number of each species of spin $<n_{i \alpha}>=\frac{1}{2}$, we find that with $J=\frac{4t^2}{U}$,  the energy of one particle can be related with the energy per site in the antiferromagnetic chain  by the relation
 \begin{equation}
 t \cos k_0=\frac{4t^2}{U} \frac{3}{4}\cos\theta
 \end{equation}
 For $t=U$, we find
 \begin{equation}
\cos\theta =\frac{1}{3}\cos k_0
 \end{equation}
 So the concurrence for different values of $k_0$ at $t=U$ can be obtained in terms of the Berry phase acquired by the spin through the relation
 \begin{equation}
 C=\frac{1}{2}(1+|\cos\theta|)_{\theta \neq 0,\pi}=\frac{1}{2}(1+\frac{1}{3}|\cos k_0|)_{k_0 \neq 0,\pi}
 \end{equation}

From this, we can have a numerical estimate of concurrence for different values of $k_0$. In fact, we find for $k_0 = \pi/4, \pi/2, 3\pi/4$,
$C=  .62, .5, .62$ respectively. Again, from eqn.(26) we note that for $k_0=0,\pi$ we get $C=1$. It is found that the results are in  good agreement with the values of concurrence obtained by Ramsak et.al. \cite{8} from an analysis of the spin flip and nonflip amplitudes of two electron interaction for wavepackets with well defined momentum.
\section{Summary and Conclusion}

To summarize, the present analysis shows that the spin entanglement of two identical fermions at two different spatial regions can be described by the Berry phase acquired by the spins in the two domains. We have considered two identical fermions, localised in two different spatial regions whose spins interact through magnetic coupling. As the study of entanglement in the solid state environment is important,  to substantiate our derivation, we have considered two electrons in two different spatial regions in Hubbard model.  We have derived the concurrence for their spin entanglement in terms of the Berry phase acquired by their spins. We have found that the results obtained in our method (value of the concurrence in Hubbard model) are in good agreemnent with the existing results in the literature \cite{8}. 

We may conclude by mentioning that it is difficult \cite{h1,h2,h3,h4,h5} to have any directly measurable observable which corresponds to entanglement of a given arbitrary quantum state. In this novel approach, the value of concurrence, which is a degree of measure to quantify spin entanglement of two fermions, can be estimated by the observed Berry phase acquired  by their spins. Furthermore,  as we have already shown that the concurrence for the entanglement of distinguishable spins in a spin system can be related to the Berry phase acquired by their spins\cite{1}, the present approach generalises the relationship between entanglement of  two distinguishable  and indistinguishable fermions.

{\it Acknowledgement:} We are thankful to the referees for their constructive comments and good suggestions.
  

\begin{thebibliography}{*}
\bibitem{sch} B. Schumacher, Phys. Rev. A {\bf 54}, 2614 (1996)
\bibitem{sc} S. Lloyd, Science {\bf 261}, 1589 (1993); D. P. Vincenzo, Scienc {\bf 270}, 255 (1995)
\bibitem{ek} A. Ekert, Phys. Rev. Lett. {\bf 67}, 661 (1991)
\bibitem{ben} C. H. Bennet et. al., Phys. Rev. Lett.{\bf 70}, 1095 (1993)
\bibitem{a1} C. Bennet, H. Bernstein, S. Popescu and B. Schumacher, Phys. Rev. A {\bf 53}, 2046 (1996)
\bibitem{b1} S. Hill and W. K. Wootters, Phys. Rev. Lett. {\bf 78}, 5022 (1997)
\bibitem {c1} V. Vedral, M. B. Plenio, M. A. Rippin and P.L. Knight, Phys. Rev. Lett. {\bf 78}, 2225 (1997)
\bibitem{d1} W.K. Wootters, Phys. Rev. Lett. {\bf 80}, 2245 (1998)
\bibitem{3} G. C. Ghirardi and L. Marinatto, Phys. Rev. A {\bf 70}, 012109 (2004)
 \bibitem{4} K Eckert, J. Schliemann, G. Bruss and M. Lewenstein, Ann. Phys. {\bf 299}, 88 (2002)
 \bibitem{5}J R Gittings and A. J. Fisher, Phys. Rev. A {\bf 66}, 032305 (2002)
 \bibitem{6} Z. Huang and S. Kais, Chem. Phys. Lett. {\bf 413}, 1, (2005)
 \bibitem{7} F. Buscemi, P. Bordone and A. Bertoni, Phys. Rev. A {\bf 73}, 052312 (2006)
  \bibitem{2} V. Vedral, Central Eur. J. Phys. {\bf 2}, 289 (2003)
  \bibitem{8}A Ramsak, I.Sega and J. A. Jefferson, Phys. Rev. A {\bf 74}, 010304(R)(2006)
\bibitem{1} B.Basu, Europhys. Lett. {\bf 73}, 833 (2006);
 B. Basu and P. Bandyopadhyay, Int. J.Geo. Meth. Mod.Phys. {\bf 4}, No. 5 707 (2007)
 %\bibitem{9} E H Lieb and F.Y. Wu, Phys. Rev. Lett. {\bf 25}, 543 (1968)
 \bibitem{10} R. A. Bertlmann, K. Durstberger, Y. Hasegawa,B. C. Hersmayer,
 Phys. Rev. A {\bf 69}, 032112 (2004)
 \bibitem{11} J. Hubbard, Proc. Roy. Soc.(London)A {\bf 276}, 238 (1963)
 \bibitem{h1} A. Peres, Phys. Rev. Lett.{\bf 77}, 1413 (1996)
 \bibitem{h2} G. Vidal and R. F. Werner, Phys. Rev. A{\bf 65}, 032314 (2002)
 \bibitem{h3} A. G. White, D.F.V. James, P.H.Eberhe et.al., Phys. Rev. Lett.{\bf 83}, 3103 (1999)
 \bibitem{h4} H. Haffner et.al. Nature (London){\bf 438}, 443(2005)
 \bibitem{h5} K.J. Ruch, P. Walther and A. Zerlinger, Phys. Rev. Lett.{\bf 94}, 070402 (2005)
 
 
 

          \end{thebibliography}
          \end{document}